\documentclass{mem}
\usepackage{natbib}
\usepackage{txfonts}
\usepackage{balance}
\usepackage[pdftex]{graphicx}
\usepackage{flushend}
\usepackage[breaklinks,pdftex]{hyperref}
\idline{95}{46}
\begin{document}
\def\teff{$T\rm_{eff }$}
\def\kms{$\mathrm {km s}^{-1}$}

\title{The Ca II K index-Mg II index relation}
   \subtitle{A Hilbert-Huang Transform approach}

\author{
R. \,Reda\inst{1}
\and V. \,Penza\inst{1}
          }

\institute{
Dipartimento di Fisica, Università degli Studi di Roma "Tor Vergata", Via della Ricerca Scientifica 1, Roma, 00133, Italy
\\
\email{raffaele.reda@roma2.infn.it}\\
}

\authorrunning{R.Reda \& V. Penza}

\titlerunning{The Ca II K index-Mg II index relation}

\date{Received: Day Month Year; Accepted: Day Month Year}

\abstract{
The solar activity, which is driven by a variable magnetic field, exhibits changes along several time scales, the 11-year being the most known. In addition to the SunSpot Number, the Ca II K index and the Mg II index are indices widely employed among those proposed to quantify the solar activity, also because of their ability to trace the solar UV emission. In this work, we compare the Ca II K 0.1nm emission index to the Mg II index over the time interval 1978-2017, which covers almost four solar cycles. We show that they are strongly correlated across each solar cycle (r$\geq$0.94), providing the corresponding linear regression fit parameters. The Hilbert-Huang Transform is then used to decompose such indices into their intrinsic mode of oscillation. By studying how their components are correlated over the different time scales, it is found that the maximum correlation is observed at the 11-year scale, while the correlation is less strong going to smaller time scales.
}
\maketitle{}

\section{Introduction}
The Sun is characterized by a variable magnetic activity which displays different scales of variation both in time, from a 
few minutes to centuries, and in wavelength. To quantify the magnetic activity level a large variety of indices (or proxies) 
have been proposed since the first observations. Based on their calculation or derivation methods, these indices can be 
mainly divided into two subgroups: synthetic indices and physical indices \citep[see e.g.,][]{Usoskin2017}. Synthetic 
indices (e.g., SunSpot Number) are computed from observed data by using a prescribed procedure, while physical indices 
(e.g., spectral line proxies) directly quantify the measured value of a physical observable, such as the flux in a given 
spectral band. 
Among physical indices, which constitute the best way to quantify the solar activity, the Mg II index and the Ca II index are the most used ones. Indeed, 
they are known to be excellent solar activity proxies, as their cores form in chromospheric layers and go into strong 
emission in the presence of magnetic field \citep[e.g.,][]{Linsky2017}. In particular, the Ca II K index brightness 
intensity is strongly related to the intensity of the solar magnetic field \citep{Babcok1955, Ermolli2018}, as well as to the line-of-sight unsigned 
magnetic flux density \citep{Chatzistergos2019}, thus constituting a reliable starting point for Total Solar Irradiance (TSI) reconstructions \citep[see e.g.,][]{Chatzistergos2021, Chatzistergos2022, Penza2022}. On the other hand, the Mg II index has been proven to be an excellent proxy 
for the solar UV irradiance \citep{DudokdeWit2009} and it is also used in TSI models as chromospheric component 
\citep[see][and references therein]{Viereck2004}.
The Ca II K line has been monitored from ground for a long period, so measurements exist since the beginning of the 20th century \citep[e.g.,][]{Bertello2016}. Instead, Mg II lines observations date back to late-1970s \citep[e.g.,][]{Viereck2004}, since they can only be performed from space.
While the relationship of the Ca II K index with Mg II index has been investigated by \citet{Donnelly1994} and \citet{DeLand2013} on shorter time intervals, this has never been done over a four solar cycle time period. Moreover, the relationship has not been explored as a function of the time scale variability, which represents the main aim of this work.

\section{Data}
\label{data section}
The dataset used in this work consists of two physical solar activity proxies: the Ca II K-0.1nm emission index (referred to simply as Ca II K index, 0.1nm is the equivalent width) and the Mg II core-to-wing ratio (or simply Mg II index).
The Ca II K index accounts for the intensity of the emission peak at 393.4 nm in the solar spectrum (the Ca II K line). This line originates in the intermediate solar atmospheric layer, the chromosphere, at a height between 300 and 1200 km \citep[see e.g.,][]{Ermolli2010}. One of the most extended dataset of the Ca II K index has been provided by \citet{Bertello2016}. Such composite has been derived by cross-calibrating data from the Kodaikanal Solar Observatory (1907-2013), from the K-line monitor program at Sacramento Peak (1976-2015) and from the ISS-SOLIS facility (2006-2017), the latter two managed by the National Solar Observatory (NSO). At the end, it contains data with monthly resolution from 1907 to October 2017.
The Mg II index is defined as the ratio between the H \& K emission doublet (279.6 nm and 280.3 nm respectively) and the intensity in the line wings \citep{Heath1986}. The core of the Mg II lines originates in the chromosphere, while the wings in the photosphere \citep{Bruevich2014}. The Mg II index used here is the one from the Bremen University \citep[see][for further details]{Viereck2004}, mainly based on inter-calibrated measurements from the satellites SBUV(/2), U/SOLSTICE, GOME, SCIAMACHY, GOME-2A and GOME-2B, containing measures starting from November 1978 to date. 
\begin{figure}
    \centering
    \includegraphics[width=0.5\textwidth]{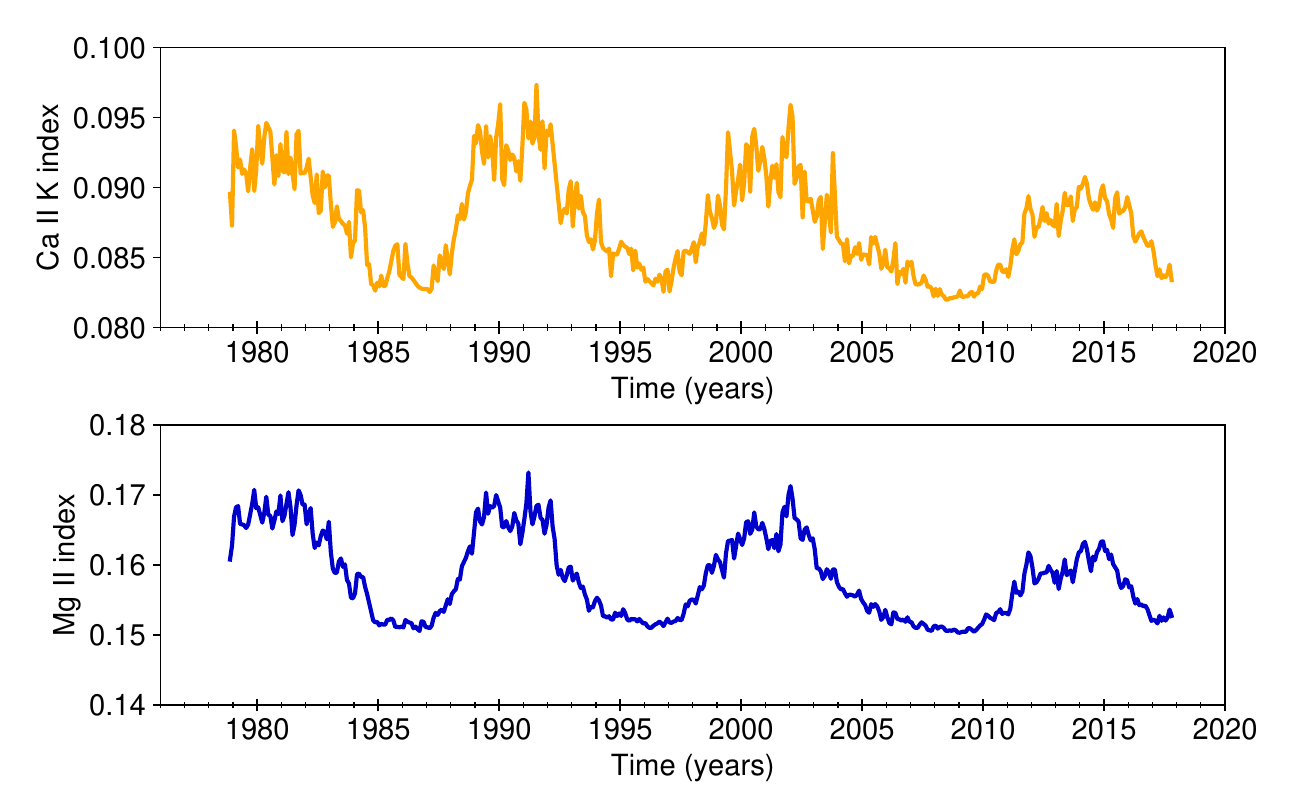}
    \caption{Monthly averages of Ca II K index (top, orange) and Mg II index (bottom, blue).}
    \label{caII_mgII}
\end{figure}

\section{Ca II K index - Mg II index relation}
\label{Ca_Mg relation section}

\subsection{Monthly averages relation}

\begin{table*}
\begin{center}
\begin{tabular}{c|ccc}
\hline
Solar cycle & Corr. coeff.  & $\alpha\pm\Delta\alpha$ & $\beta\pm\Delta\beta$  \\
\hline
21  &  0.940  &  $0.5012\pm0.0190$  &  $0.0079\pm0.0031$ \\
22  &  0.955  &  $0.5877\pm0.0169$  &  $-0.0050\pm0.0027$ \\
23  &  0.954  &  $0.5899\pm0.0158$  &  $-0.0059\pm0.0025$ \\
24  &  0.938  &  $0.6013\pm0.0216$  &  $-0.0076\pm0.0034$ \\
All  &  0.946  &  $0.5632\pm0.0090$  &  $-0.0016\pm0.0014$ \\
\hline
\end{tabular}
\end{center}
\caption{Cycle-to-cycle correlation between Ca II K index and Mg II index. Column 1: Solar cycle number; Column 2: 
Pearson's 
correlation coefficient; Columns 3 and 4: Linear regression fit parameters of the relation $Ca\;II\;K = (\alpha \pm 
\Delta\alpha)\,Mg\;II + (\beta \pm \Delta\beta)$.}
\label{cycle_to_cycle_table}
\end{table*}

\begin{figure}
    \centering
    \includegraphics[width=0.4\textwidth]{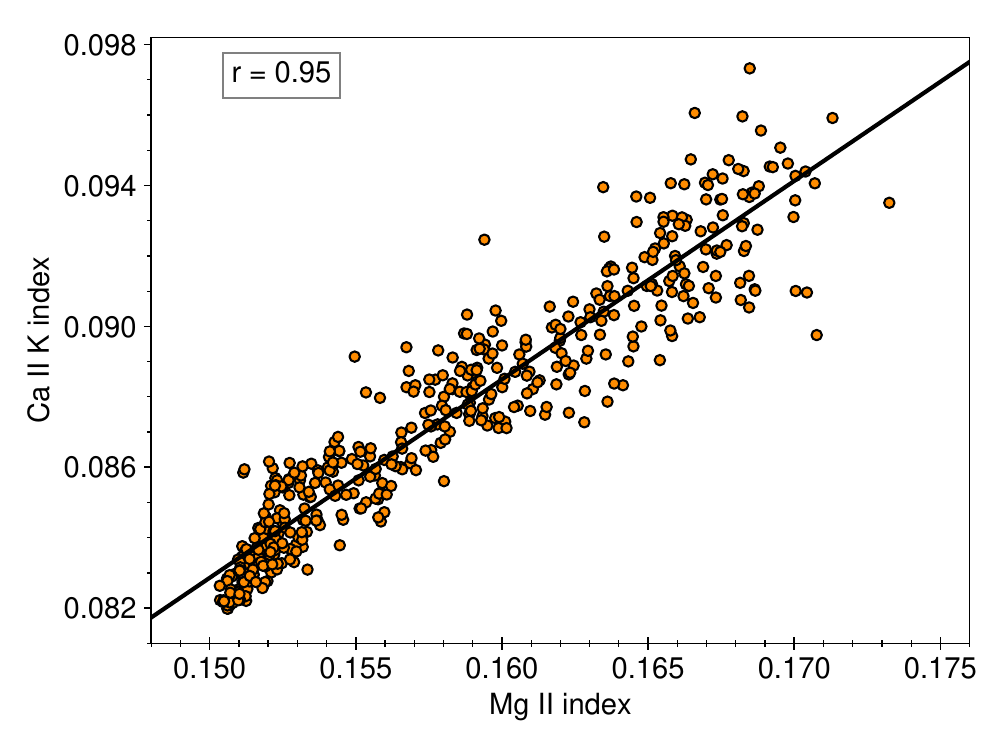}
    \caption{Scatter plot showing the correlation between the Ca II K index and the Mg II index. The black line shows the best linear fit.}
    \label{correlation}
\end{figure}

\begin{figure}
    \centering
    \includegraphics[width=0.5\textwidth]{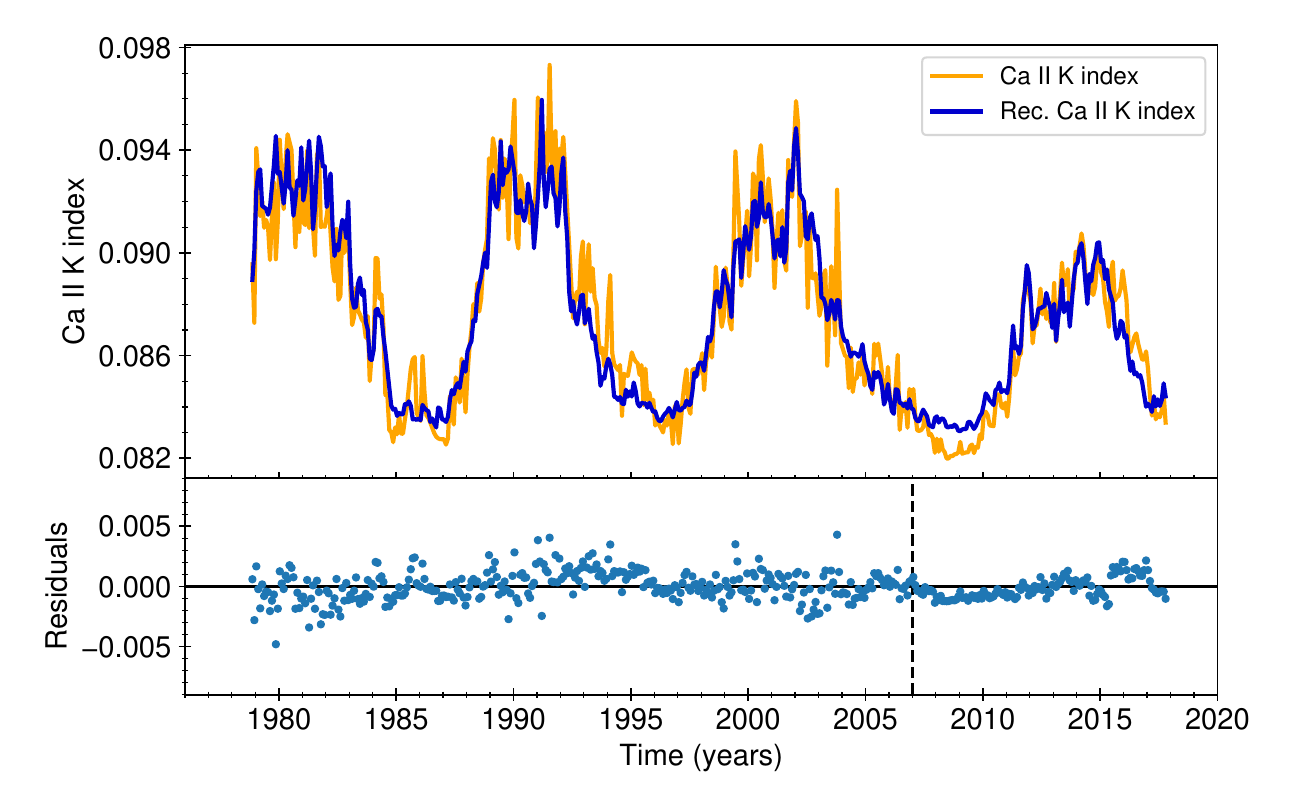}
    \caption{Comparision of the Ca II K index (orange) with the one reconstructed from the Mg II index (blue) by using the fit's parameters of the last row of Table \ref{cycle_to_cycle_table} in relation \ref{fit_ca_mg}. The residuals plot on the bottom shows the difference between the two time series.}
    \label{comparison_residuals}
\end{figure}

The relation of the Ca II K index with Mg II index is here analyzed along the overlapping time interval of the two series, which goes from November 1978 to October 2017. The two time series are shown in Figure \ref{caII_mgII}. Figure \ref{correlation} shows the scatter plot of the Mg II index with the Ca II K index. The two solar activity proxies are clearly linearly correlated, with a Pearson correlation coefficient r=0.95. This result allow to link the two proxies by means of a linear regression fit, as follows:
\begin{equation}
Ca\;II\;K = (\alpha \pm \Delta\alpha)\,Mg\;II + (\beta \pm \Delta\beta)
\label{fit_ca_mg}
\end{equation}
where $\alpha$ is the slope of the fit and $\beta$ is the intercept. The values of the fit parameters for each solar cycle, 
as well as for the whole time interval, are reported in Table \ref{cycle_to_cycle_table}, together with the corresponding 
correlation coefficients. Ca II K index and Mg II index result to be strongly correlated over each solar cycle ($r \geq 
0.94$). In addition, we notice that for solar cycle 21 the parameters of the linear regression are slightly different 
compared to those of the subsequent cycles, probably due to the fact that we miss the cycle's ascending phase data. 
Considering the whole extent of the time series, the values obtained for the fit's parameters are $\alpha=0.5632\pm0.0090$ 
and $\beta=-0.0016\pm0.0014$. The latter values can be used in Eq. \ref{fit_ca_mg} to see how well the Mg II index is able 
to reproduce the Ca II K index behaviour. Fig. \ref{comparison_residuals} shows a comparison between the Ca II K index 
\citep[][ composite]{Bertello2016} and the Ca II K index reconstructed via the Mg II index. As it can be seen, the two 
signals are in good agreement along all the phases of the solar cycle (the residuals are very small). However, it is 
possible to notice that the residuals become systematically smaller after 2007 (marked by the dashed vertical line), 
corresponding to the starting date of the ISS-SOLIS data in Ca II K composite.
Once a relation between Ca II K index and Mg II index is found, it opens the possibility to extend the former index up to present and, vice versa, to reconstruct the latter in the past.

\subsection{Time scales relation by means of Hilbert-Huang Transform}
\label{HHT subsection}

\begin{figure*}
    \centering
    \includegraphics[width=0.7\textwidth]{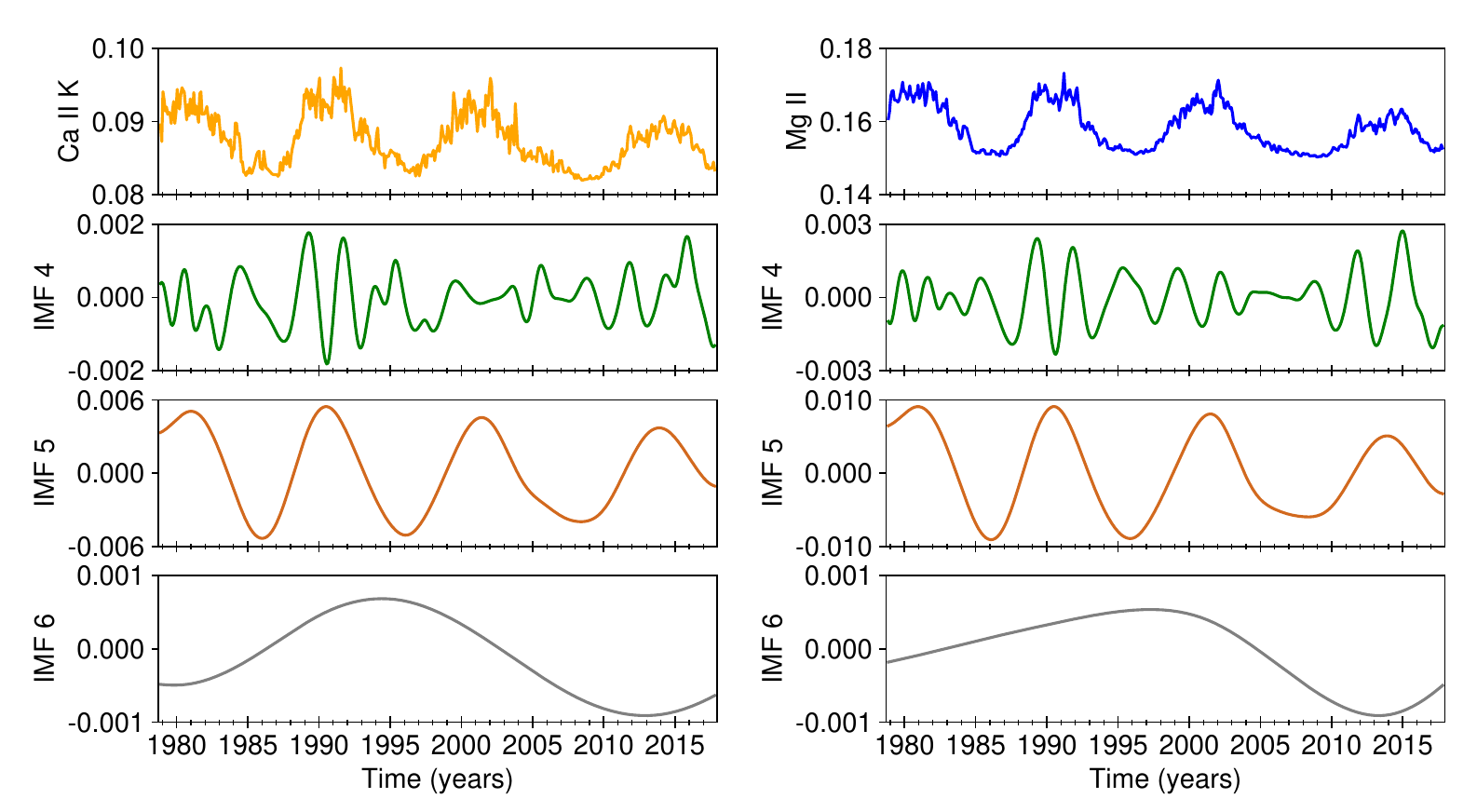}
    \caption{Empirical Mode Decomposition of Ca II K index (left panels) and Mg II index (right panels). The top row shows the starting signal, the subsequent rows show the IMFs 4-6.}
    \label{EMD_ca_mg}
\end{figure*}

The Hilbert-Huang Transform (HHT) is an extremely powerful data-driven method used in the time-frequency analysis of non-linear and non-stationary time series. It consists of two main steps:
\begin{enumerate}
    \item \textit{Empirical Mode Decomposition (EMD)} \\
    Given a time series $x(t)$, the EMD enables to write it as the sum of $n$ components, called Intrinsic Mode Functions (IMFs) or simply modes, plus a further term $res(t)$ which represents the residual of the decomposition.
    \begin{equation}
    x(t) = \sum_{i=1}^n IMF_{i}(t) + res(t)
    \end{equation}
    \item \textit{Hilbert Spectral Analysis} \\
    It consists in applying the Hilbert Transform to the set of IMFs obtained in the previous step. This results in the computation of the analytic signal of each IMF, from which it is possible to derive the instantaneous amplitude $A_i(t)$, phase $\phi_i(t)$ and frequency $f_i(t)$. In particular, from the latter quantity it is possible to obtain insights about the mean time scale of each mode, as $<t_i>=1/<f_i(t)>$. 
\end{enumerate}
A more detailed description of the two steps of the HHT can be found in \citet{Huang1998, Reda2024}. We apply here the HHT to the time series of Ca II K index and Mg II index, in order to show how it makes possible to decompose the signals into their time scale components, allowing to study the relation between each of them. Fig. \ref{EMD_ca_mg} shows some of the results of the EMD (only IMFs 4-6 are reported) for Ca II K index (left panels) and for Mg II index (right panels).
\begin{table*}
\begin{center}
\begin{tabular}{c|ccc}
\hline
IMF \# & Ca II K &  Mg II  & Corr. Coeff \\
  &  $\mathrm{<t>}$ (yrs)  & $\mathrm{<t>}$ (yrs) &  (Ca II K - Mg II) \\
\hline
1  &  0.41 &  0.38 &  0.47 \\
2  &  0.64 &  0.67 &  0.70 \\
3  &  1.48 &  1.63 &  0.70 \\
4  &  3.01 &  3.04 &  0.73 \\
5  &  11.36 &  11.35 &  0.99 \\
6  &  36.91 &  40.31 &  0.95 \\
\hline
\end{tabular}
\end{center}
\caption{Relation between the Ca II K index and Mg II index modes. Column 1: IMF number; Columns 2 and 3: Characteristic (or mean) time scale of the Ca II K index and Mg II index modes, respectively; Column 4: Pearson's correlation coefficient.}
\label{IMFs_correlation_table}
\end{table*}
In Table \ref{IMFs_correlation_table} we also report the characteristic time scales of the IMFs of Ca II K index and Mg II Index. For both signals, the IMF 5 is the one corresponding to the 11-year solar cycle and also the one with the highest weighted variance (77\% for Ca II K and 87\% for Mg II), being the component accounting for the most prominent variations. The advantage of having decomposed the two signals into the intrinsic modes of oscillation is that it becomes now possible to study their relation on the different time scales. To this scope, the last column of Table \ref{IMFs_correlation_table} contains the correlation coefficients of the Ca II K index and Mg II index IMFs. It is possible to notice that while the correlation is at least moderate ($\geq$ 0.45) across all the time scales, it becomes higher moving to larger time scales. The maximum correlation is observed between the IMFs 5, hence at the solar cycle scale.

\section{Conclusions}
\label{conclusion section}
The relation between the Ca II K index and Mg II index has been here studied. Although both emission lines originate in the solar chromosphere, the different height where they form can give rise to intensity differences. By analyzing the cycle-to-cycle correlation, we find that Ca II K index and Mg II index exhibit a strong correlation over each solar cycle, as well as over the whole time interval investigated. The linear relation \ref{fit_ca_mg}, together with the regression fit parameters provided in Table \ref{cycle_to_cycle_table}, can be very useful to fill gaps in the Ca II K index time series, as well as to easily reconstruct Ca II K data further than October 2017, as done by \citet{Reda2023MNRAS}. Since the Mg II index is freely accessible to date, such a relation can in principle be used to obtain Ca II K index to present. Conversely, it is possible to use the Ca II K in order to estimate the Mg II index trend back in the past. 
Furthermore, the use of Hilbert-Huang Transform to decompose Ca II K index and Mg II index enables a more in-depth exploration of the relationship between these indices in terms of their variations across different time scales. In this respect, it is here found that the maximum correlation occurs on the 11-year time scale, while on smaller ones the correlation is less strong. A more detailed analysis will be provided in a forthcoming article.

\begin{acknowledgements}
R.R. acknowledges the support from the European Union’s Horizon 2020 research and innovation program under grant agreement No. 824135 (SOLARNET). The authors thank Luca Giovannelli for the useful discussion.
\end{acknowledgements}

\bibliographystyle{aa}
\bibliography{bibliography}

\end{document}